\documentclass[twocolumn,showpacs,preprintnumbers,amsmath,amssymb,groupedaddress,superscriptaddress]{revtex4}
\usepackage{graphicx}
\usepackage{dcolumn}
\usepackage{bm}

\begin{document}


\title{A novel absorption resonance for all-optical atomic clocks}
\author{Sergei Zibrov}
\affiliation{Department of Physics, Harvard University, Cambridge,
Massachusetts, 02138} \affiliation{Moscow State Engineering
Physics Institute, Moscow, 115409, Russia}
 \affiliation{Lebedev Institute of Physics, Moscow, 117924,
Russia}
 \affiliation{Harvard-Smithsonian Center for Astrophysics,
Cambridge, Massachusetts, 02138}
\author{Irina Novikova}
\affiliation{Harvard-Smithsonian Center for Astrophysics,
Cambridge, Massachusetts, 02138}
\author{David F. Phillips}
\affiliation{Harvard-Smithsonian Center for Astrophysics,
Cambridge, Massachusetts, 02138}

\author{Aleksei V. Taichenachev}
\author{Valeriy I. Yudin}
\affiliation{Institute of Laser Physics SB RAS and Novosibirsk
State University, Novosibirsk, 630090, Russia}

\author{Ronald L. Walsworth}
\affiliation{Department of Physics, Harvard University, Cambridge,
Massachusetts, 02138} \affiliation{Harvard-Smithsonian Center for
Astrophysics, Cambridge, Massachusetts, 02138}
\author{Alexander S. Zibrov}
 \affiliation{Department of Physics, Harvard University, Cambridge, Massachusetts, 02138}
\affiliation{Lebedev Institute of Physics, Moscow, 117924, Russia}
 \affiliation{Harvard-Smithsonian Center for Astrophysics,
Cambridge, Massachusetts, 02138}

\date{\today}

\begin{abstract}
  We report an experimental study of an all-optical three-photon-absorption resonance (known
  as a ``\emph{N}-resonance'') and discuss its potential application as an
  alternative to atomic clocks based on coherent population trapping
  (CPT).  We present measurements of the \emph{N}-resonance contrast, width and light-shift for the $D_1$ line of
  ${}^{87}$Rb with varying buffer gases, and find good agreement with an analytical model of this novel resonance.
  The results suggest that \emph{N}-resonances are
  promising for atomic clock applications.
\end{abstract}

%
%

\maketitle

There is great current interest in developing compact, robust
atomic clocks with low power consumption and fractional frequency
stability better than $10^{-12}$ for a wide variety of
applications. In recent years, significant progress toward this
goal has been achieved using coherent population trapping (CPT)
resonances in atomic vapor \cite{Cyr,Leo,Vanier03,Finland}.  In
this paper, we investigate an all-optical three-photon-absorption
resonance in Rb vapor~\cite{Zibrov-Observation}, known as an
``\emph{N}-resonance'', which combines advantages of CPT and
traditional optically-pumped double-resonance. We find that the
\emph{N}-resonance provides high contrast with modest systematic
frequency shifts, and thus may be suitable for small, stable
atomic clocks.

\begin{figure}
\includegraphics[width=1.0\columnwidth]{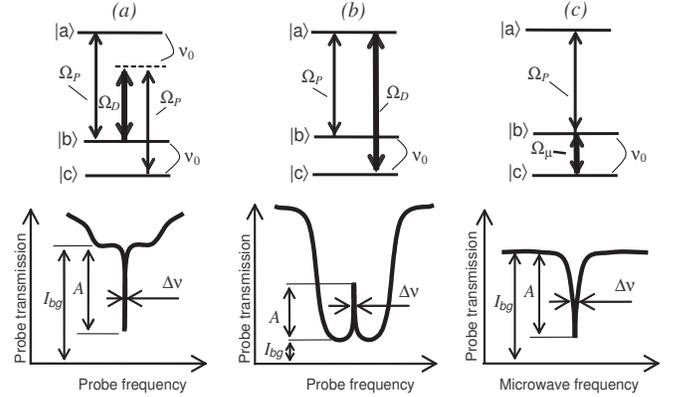}%
\caption{ Level diagrams (top row) and schematic representations
of probe light transmission spectra for \emph{(a)}
\emph{N}-resonance, \emph{(b)} CPT resonance, and \emph{(c)}
 optical-pumping double-resonance schemes. Shown in the level diagrams are
 the relevant probe ($\Omega_P$), drive ($\Omega_D$) and microwave ($\Omega_{\mu}$)
 fields, as well as the ground-state hyperfine splitting $\nu_0$. Shown in the
 schematic
 spectra are the full width ($\Delta\nu$)
 and
relative intensity ($A$) of the clock resonance, and
 the intensity of the background transmitted light ($I_{bg}$).
   }
\label{fig1}
\end{figure}

An \emph{N}-resonance is a three-photon, two-field resonance, as
shown in Fig.~\ref{fig1}a. An optical probe field, $\Omega_P$,
resonant with the transition between the higher-energy hyperfine
level of the ground electronic state ($|b\rangle$) and an excited
state ($|a\rangle$), optically pumps the atoms into the lower
hyperfine level ($|c\rangle$), leading to increased transmission
of the probe field through the medium. A second, off-resonant
optical drive field, $\Omega_D$, is detuned to lower frequencies
than the $|b\rangle\rightarrow|a\rangle$ transition.  If the
difference frequency between $\Omega_P$ and $\Omega_D$ is equal to
the hyperfine frequency, a two-photon resonance is created,
driving atoms coherently from state $|c\rangle$ to $|b\rangle$,
followed by a one-photon absorption from field $\Omega_P$ which
drives the atoms to the excited state $|a\rangle$
~\cite{happer_note}. Thus, the absorption spectrum of the
$\Omega_{P}$ field will have two distinct features (Fig.\
\ref{fig1}a, bottom row): a broad Doppler background due to linear
absorption and a narrow resonance because of the three-photon
nonlinear process. Importantly for clock applications, the
\emph{N}-resonance is all-optical; also, nearly 100\%  of the
probe field is absorbed on resonance, which greatly reduces the
practical effects of both shot noise, as well as phase noise due
to frequency/intensity noise conversion in the optical
fields~\cite{camparo97,Leo}.

For comparison, we also plot in Fig.\ \ref{fig1} typical level
diagrams and schematic spectra of the probe light transmission for
CPT and traditional optically-pumped double-resonance schemes. CPT
is a two-photon transmission resonance (Fig.~\ref{fig1}b) in which
a coherence between two hyperfine levels is created by two
resonant fields ($\Omega_P$ and $\Omega_D$) whose frequency
difference is equal to the hyperfine frequency.  The absorption
for both optical fields decreases due to destructive interference
of the absorption amplitudes, and a narrow transmission peak is
observed. Several groups~\cite{Finland,Leo} have achieved
fractional hyperfine frequency stabilities below $10^{-12}$ with
CPT-based clocks, which are also promising for miniaturization.
In the traditional optically-pumped double-resonance clock
(Fig.~\ref{fig1}c), one optical field (from a lamp or laser diode)
is resonant with one of the allowed transitions
($|b\rangle\rightarrow |a\rangle$), and thus optically pumps atoms
to the other hyperfine sublevel ($|c\rangle$).  A microwave field
 resonant with the ground-state hyperfine
transition is applied, thereby redistributing the populations
between the hyperfine levels and leading to a narrow dip in the
transmission spectrum of the optical field.
The width of this absorption feature is determined by the
intensities of both the optical and microwave fields as well as
the atoms' hyperfine decoherence rate. With careful optimization
of operational parameters, short-term fractional stabilities of
$10^{-11}$ may be achieved~\cite{Vanier-book}.

In practice, the frequency stability of an atomic clock limited by
photon shot noise is given by the Allan deviation, $\sigma(\tau)$,
as~\cite{Vanier03}:
\begin{equation}
  \sigma(\tau)=\frac{1}{4}\sqrt{\frac{\eta
  e}{I_{bg}}}\frac{1}{\nu_0} \frac{\Delta\nu}{C}\tau^{-1/2}
\label{e.stability}
\end{equation}
where $\nu_0$ is the atomic reference frequency, $\Delta\nu$ is
the full width of the resonance,
$e$ is the electron charge, $\eta$ is the photodetector
sensitivity (measured optical energy per photoelectron)
and $\tau$ is the integration time.
The resonance contrast $C\equiv A/I_{bg}$, where $A$ is the
relative intensity of the clock resonance and $I_{bg}$ is the
intensity of the background transmitted light (adopting notation
similar to reference~\cite{Vanier03}).  These contrast parameters
are shown graphically in Fig.~\ref{fig1} for the
\emph{N}-resonance, CPT resonance, and optically-pumped
double-resonance schemes.
%
\begin{figure}
\includegraphics[width=0.9\columnwidth]{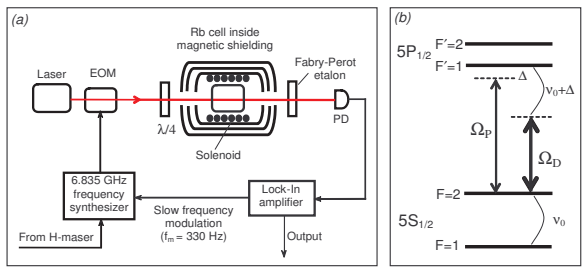}%
\caption {
  (a) Experimental apparatus. (b) Energy levels and applied fields
  ($\Omega_P$, $\Omega_D$) for \emph{N}-resonances on the $D_1$ line
  of $^{87}$Rb. $\nu_0$ is the ground electronic state hyperfine
  splitting and $\Delta$ is the detuning of the probe field from the
  $F=2\rightarrow F^\prime=1,2$ resonance. }
\label{fig2}
\end{figure}

Figure~\ref{fig2}a shows a schematic of our \emph{N}-resonance
experimental set-up.  We derived the probe and drive optical
fields ($\Omega_P$ and $\Omega_D$) by phase modulating the output
of an external cavity diode laser ($\approx 12$~mW total power)
tuned in the vicinity of the $D_1$ line of Rb
($5^2S_{1/2}\rightarrow5^2P_{1/2}$, $\lambda \simeq 795$~nm). An
electro-optic modulator (EOM) produced the phase modulation of the
optical field at a frequency near the ground electronic state
hyperfine frequency of ${}^{87}$Rb ($\nu_0\simeq 6.835$~GHz).
Approximately 2\% of the incident laser power was transferred to
each first-order sideband, with the remainder residing in the
carrier. The laser beam was then circularly polarized using a
quarter wave plate and weakly focused to a diameter of 0.8 mm
before entering the Rb vapor cell.

We employed Pyrex cylindrical cells containing isotopically
enriched ${}^{87}$Rb and either 40~Torr Ne buffer gas or a 10~Torr
Ne + 15~Torr Ar mixture. During experiments, the vapor cell under
study was heated to 55\,${}^\circ$C using a blown-air oven. The
cell was isolated from external magnetic fields with three layers
of high permeability shielding. A small ($\approx 10$~mG)
longitudinal magnetic field was applied to lift the degeneracy of
the Zeeman sublevels and separate the desired $F=1$, $m_F=0$ to
$F=2$, $m_F=0$ clock transition (no first-order magnetic field
dependence) from the $m_F=\pm1$ transitions (first-order Zeeman
splitting).

To produce the \emph{N}-resonance we tuned the high frequency
optical sideband (serving as the probe field $\Omega_P$) near
resonance with the $5S_{1/2}\ F=2\rightarrow 5P_{1/2}\
F^\prime=1,2$ transitions.
The strong laser carrier field at a frequency 6.835~GHz below this
transition was used as the drive field $\Omega_D$ (see Fig.\
\ref{fig1}a). Note that we operate in the regime of relatively low
laser power and atomic density, which is different from
\cite{Zibrov-Observation}. In the present case all four-wave
mixing processes are insignificant, and the far-off-resonance
lower frequency sideband had negligible effect on the atoms.
The strong drive field and the lower sideband were filtered from
the light transmitted through the cell using a quartz, narrow-band
Fabry-Perot etalon (FSR~=~20~GHz, finesse~=~30), tuned to the
frequency of the probe field and placed before the photodetector.
Such selective detection reduces the light background ($I_{bg}$)
by eliminating nonreasonant leakage from the drive field; and also
increases the absorption amplitude ($A$) by eliminating the
stimulated Raman drive field created at two-photon resonance
~\cite{Zibrov-Observation}. Our analytical modelling of the
\emph{N}-resonance -- based on the method developed in
~\cite{taichenachev03pra,knappe02apb} for CPT resonances --
indicates that the absorption amplitude $A$ increases by $\approx
1.7$ when only the probe field transmission is detected. (Details
of this analytical model will be described in a future
publication).

\begin{figure}
\includegraphics[width=0.9\columnwidth]{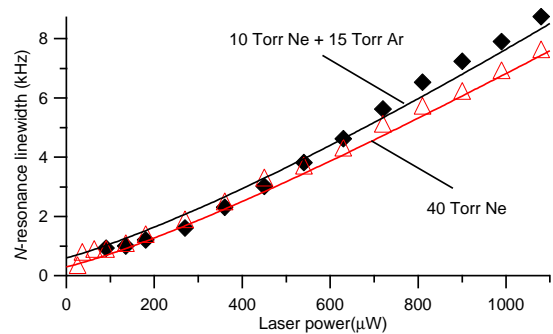}
\caption {
  Measured \emph{N}-resonance linewidth as a function of total
  incident   laser power for ${}^{87}$Rb vapor cells filled with
  40 Torr of Ne buffer gas ($\triangle$) and a mixture of $10$~Torr
  Ne and $15$~Torr Ar ($\blacklozenge$). The probe field
  is tuned $\approx 300$~MHz below the $F=2\rightarrow F^\prime=2$ transition of
  ${}^{87}$Rb. Solid lines are calculated linewidths using an analytical \emph{N}-resonance model.}
\label{fig3}
\end{figure}

Fig.~\ref{fig3} shows measured \emph{N}-resonance linewidths for
two different buffer gases.
%
%
At lower laser power, linewidths $<1$~kHz are observed: e.g.,
$\Delta \nu \approx 300$~Hz  at 50 $\mu$W total laser power for
the 40 Torr Ne cell. At larger laser powers the linewidth
increases approximately linearly with laser power. As also shown
in Fig.~\ref{fig3}, our calculations are in good agreement with
the measured variation of linewidth with laser power.

\begin{figure}
\includegraphics[width=1.0\columnwidth]{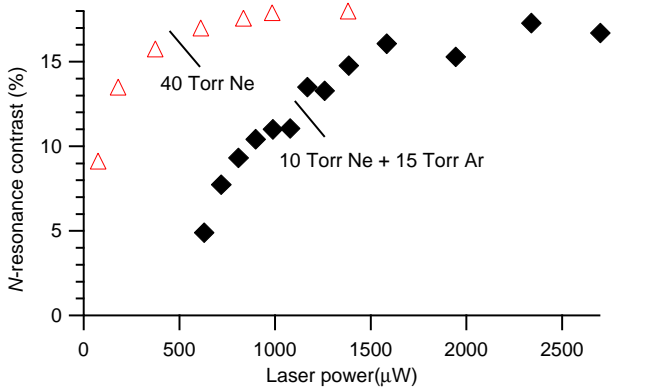}
\caption {
 Measured dependence of \emph{N}-resonance contrast
 on total incident laser power for two buffer gas cells:
 40~Torr Ne ($\triangle$) and 10~Torr Ne +
 15~Torr Ar ($\blacklozenge$). The probe field is tuned $\approx 300$~MHz below
 the $F=2\rightarrow F^\prime=2$ transition of ${}^{87}$Rb. }
\label{fig4}
\end{figure}

Fig.~\ref{fig4} shows measurements of the \emph{N}-resonance
contrast $C=A/I_{bg}$ for two buffer gas cells. For both cells the
contrast increases rapidly with laser power, and then saturates at
$C> 15\%$ for total incident laser power $\sim 1$~mW. This
saturated \emph{N}-resonance contrast exceeds the contrast that
has been achieved with CPT resonances, $C<4\%$~\cite{Vanier03}.
However, the relatively large laser power required to saturate the
\emph{N}-resonance contrast leads to an increased linewidth (see
Fig.~\ref{fig3}).
To account for these competing effects of laser power, we follow
Vanier \textit{et al.} \cite{Vanier03} and employ a resonance
quality factor $q\equiv C/\Delta\nu$ as a figure of merit for
\emph{N}-resonance clocks.
For example, for 120 $\mu$W of laser power for the $40$~Torr Ne
cell, we find $\Delta\nu \approx 1$~kHz and $C\approx 0.1$ for the
${}^{87}$Rb \emph{N}-resonance (see Figs.~\ref{fig3} and
\ref{fig4}), implying $q\approx 10^{-4}$ and an estimated
frequency stability from Eq.~(\ref{e.stability}) of
$\sigma(\tau)\sim10^{-14}{\tau}^{1/2}$.

\begin{figure}
\includegraphics[width=0.9\columnwidth,angle=0]{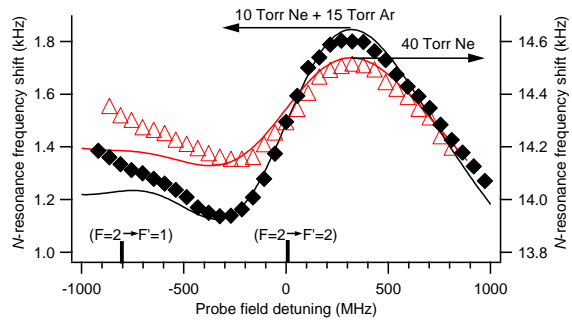}
 \caption {
   Measured variation of the \emph{N}-resonance frequency as a function of
   probe field detuning from the ${}^{87}$Rb $F=2\rightarrow F^\prime=2$ transition
   with total incident laser power of 450~$\mu$W. Right
   vertical axis: 40 Torr Ne ($\triangle$). Left vertical axis:
   25~Torr Ne-Ar mixture ($\blacklozenge$). These \emph{N}-resonance
   frequencies include a buffer gas pressure
   shift of the ${}^{87}$Rb ground state hyperfine frequency: about
    14~kHz for the 40 Torr Ne cell and 1~kHz for the 25
   Torr Ne-Ar mixture
   \protect\cite{Ottinger,Erhard}. Solid lines are calculations from an analytical model. }
\label{fig5}
\end{figure}

Importantly, Fig.~\ref{fig4} also shows that the
\emph{N}-resonance contrast reaches its maximum at lower laser
powers for the higher-pressure $40$~Torr Ne vapor cell than for
the cell with the 25 Torr Ne-Ar mixture. We attribute this
difference to slower Rb diffusion out of the laser fields at
higher buffer gas pressure, and hence reduced ground-state
coherence loss and more efficient optical pumping. In addition, we
did not observe a deterioration of the \emph{N}-resonance contrast
with increased buffer gas pressure, as has been observed for CPT
resonances \cite{Vanier03}.
%
This observation suggests that the \emph{N}-resonance may be a
good candidate for miniature atomic clocks, where high buffer gas
pressure is required to prevent rapid atomic decoherence due to
collisions with the walls of a small vapor cell.


We also characterized the light-shift of the ${}^{87}$Rb
\emph{N}-resonance. A light-shift is  a relative AC Stark shift of
atomic levels that depends on both the optical field frequency and
intensity~\cite{Mathur}. Light-shifts are a primary systematic
effect limiting the frequency stability of optically-pumped atomic
clocks and should be present for \emph{N}-resonances at some
level. To leading order in a simple two-level picture, the
light-shift, $\Delta_\mathrm{ls}$, of the clock frequency is given
by
\begin{equation}\label{eq3}
 \Delta_\mathrm{ls}=-\frac{\Delta}{\gamma_{ab}^{2}+4\Delta^{2}}|\Omega_P|^2,
\end{equation}
where $\Delta$ is the detuning of the probe field from the atomic
transition. For small $\Delta$, the light-shift is linear in the
laser frequency ($\Delta$) and intensity ($\propto |\Omega_P|^2$).
Fluctuations in these parameters are thus directly transferred to
the clock frequency.
For example, light-shifts limit the fractional frequency stability
of optically-pumped double-resonance clocks at the level of
$10^{-11}$~\cite{Budkin,Camparo,Drullinger}.

In CPT clocks, light-shifts may be eliminated, in principle. In
practice, however, the diode laser typically used in a CPT clock
is driven with strong current-modulation to produce the two
strong, resonant optical fields $\Omega_P$ and $\Omega_D$.
This modulation scheme necessarily leads to: (i) higher-order
sidebands, which, even when optimally adjusted, can induce
non-trivial second-order light-shifts \cite{Vanier98}; and (ii)
%
%
unwanted amplitude modulation of the optical fields, resulting in
an imbalance between sideband intensities as large as
10\%~\cite{Leo}.
These imperfections lead to residual light-shifts of $\sim
0.2$~Hz/($\mu$W/cm${}^2$)~(for shifts induced by changes in the
carrier field intensity) and 1 Hz/MHz~(for shifts induced by
changes in the carrier field
frequency)~\cite{Leo,Vanier03,Finland}. In practice, the short-
and medium-term frequency stability of CPT clocks is limited by
such light-shifts~\cite{Finland,Vanier03}.

\begin{figure}
  \includegraphics[width=0.9\columnwidth]{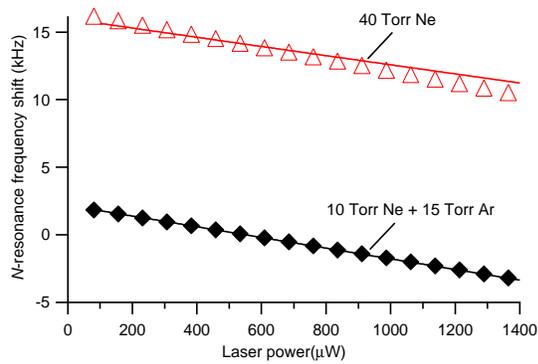}
 \caption {
   Measured dependence of the \emph{N}-resonance frequency on the total
   incident laser power, with the probe
   field tuned $\approx 300$~MHz below the $F=2\rightarrow F^\prime=2$ transition of
   ${}^{87}$Rb. Upper
   trace: 40 Torr Ne ($\triangle$). Lower trace:
   25 Torr Ne-Ar mixture ($\blacklozenge$). As in Fig.~\protect\ref{fig5},
   the \emph{N}-resonance frequencies include
   the buffer gas pressure shift. Solid lines are calculations from an analytical model.}
\label{fig6}
\end{figure}

As shown in Fig.~\ref{fig5}, we measured two extrema in the
\emph{N}-resonance light-shift as a function of probe field
frequency. (We determined the \emph{N}-resonance frequency from
the difference frequency between the probe and drive fields at
which the transmitted probe light intensity was minimized.) At
these extrema the light-shift depends quadratically on the probe
field detuning.
Additionally, the second-order light-shift near the extrema is
reduced at higher buffer gas pressure, from approximately 4.0
mHz/MHz${}^2$ for the 25 Torr Ne-Ar mixture to 2.5 mHz/MHz${}^2$
for the 40 Torr Ne cell, suggesting again that \emph{N}-resonances
may be well suited to small vapor cells employing high buffer gas
pressure.
Fig.~\ref{fig6} shows the measured dependence of the
\emph{N}-resonance light-shift on total laser power. We find a
linear dependence, with a similar variation of
25~mHz/($\mu$W/cm${}^2$) for different buffer gases. These
\emph{N}-resonance light-shifts are about an order of magnitude
smaller than for existing CPT clocks.

Finally, we note that \emph{N}-resonances on the $D_2$ line of
alkali vapor may also be promising for clock applications. Our
analytical model suggests higher \emph{N}-resonance contrast on
the $D_2$ transition due to strong collisional mixing of the
Zeeman levels in the electronic excited state, which suppresses
optical pumping to the end Zeeman levels in the ground electronic
state. (Note, CPT contrast is smaller for the $D_2$ line than for
the $D_1$ line due to pressure broadening of the excited state
hyperfine levels ~\cite{stahler02}.) Currently, diode lasers on
the $D_2$ line of Rb and Cs are more easily obtained.

%

In summary, we measured the properties of an \emph{N}-resonance on
the $D_1$ line in Rb vapor cells with varying buffer gases. We
found that this \emph{N}-resonance has greater contrast than the
corresponding CPT resonance and order-of-magnitude smaller
light-shifts. These results suggest that an all-optical atomic
clock locked to an \emph{N}-resonance may provide improved short
and medium term frequency stability compared to CPT clocks. In
addition, we found that the \emph{N}-resonance contrast does not
degrade, nor the light-shifts worsen, with increased buffer gas
pressure. Hence, \emph{N}-resonances may be good candidates for
miniature atomic clocks.

The authors are grateful to M.D. Lukin,
and V.L. Velichansky for useful discussions.  This work was
supported by DARPA.  Work at the Center for Astrophysics was
supported by ONR and the Smithsonian Institution. A.\ V.\ T.\ and
V.\ I.\ Y.\ acknowledge support from RFBR (grants no. 05-02-17086
and 04-02-16488).

%


\end{document}